\documentstyle[12pt]{article}
\catcode`\@=11

\@addtoreset{equation}{section}

\def\@normalsize{\@setsize\normalsize{15pt}\xiipt\@xiipt
\abovedisplayskip 14pt plus3pt minus3pt%
\belowdisplayskip \abovedisplayskip
\abovedisplayshortskip  \z@ plus3pt%
\belowdisplayshortskip  7pt plus3.5pt minus0pt}
\def\small{\@setsize\small{13.6pt}\xipt\@xipt
\abovedisplayskip 13pt plus3pt minus3pt%
\belowdisplayskip \abovedisplayskip
\abovedisplayshortskip  \z@ plus3pt%
\belowdisplayshortskip  7pt plus3.5pt minus0pt
\def\@listi{\parsep 4.5pt plus 2pt minus 1pt
            \itemsep \parsep
            \topsep 9pt plus 3pt minus 3pt}}

\def\underline#1{\relax\ifmmode\@@underline#1\else
        $\@@underline{\hbox{#1}}$\relax\fi}
\@twosidetrue
\relax

\catcode`@=12

\catcode`\@=11

\def\section{\@startsection{section}{1}{\z@}{3.5ex plus 1ex minus
   .2ex}{2.3ex plus .2ex}{\large\bf}}

\def\ps@headings{\def\@oddfoot{}\def\@evenfoot{}
\def\@oddhead{\hbox{}\hfill
        \makebox[.5\textwidth]{\raggedright\ignorespaces --\thepage{}--
        \hfill }}
\def\@evenhead{\@oddhead}
\def\subsectionmark##1{\markboth{##1}{}}
}

\ps@headings

\catcode`\@=12

\relax

\def\figcap{\section*{Figure Captions\markboth
        {FIGURECAPTIONS}{FIGURECAPTIONS}}\list
        {Fig. \arabic{enumi}:\hfill}{\settowidth\labelwidth{Fig. 999:}
        \leftmargin\labelwidth
        \advance\leftmargin\labelsep\usecounter{enumi}}}
 \relax
\def\tablecap{\section*{Table Captions\markboth
        {TABLECAPTIONS}{TABLECAPTIONS}}\list
        {Table \arabic{enumi}:\hfill}{\settowidth\labelwidth{Table 999:}
        \leftmargin\labelwidth
        \advance\leftmargin\labelsep\usecounter{enumi}}}
 \relax
\def\reflist{\section*{References\markboth
        {REFLIST}{REFLIST}}\list
        {[\arabic{enumi}]\hfill}{\settowidth\labelwidth{[999]}
        \leftmargin\labelwidth
        \advance\leftmargin\labelsep\usecounter{enumi}}}
 \relax

\catcode`\@=11

\def\ps@headings{\def\@oddfoot{}\def\@evenfoot{}
\def\@oddhead{\hbox{}\hfill
        \makebox[.5\textwidth]{\raggedright\ignorespaces --\thepage{}--
        \hfill }}
\def\@evenhead{\@oddhead}
\def\subsectionmark##1{\markboth{##1}{}}
}

\ps@headings

\relax

\def\firstpage#1#2#3#4#5#6{
\begin{document}
\begin{titlepage}
\nopagebreak
\title{\begin{flushright}
        \vspace*{-1.5in}
        {\normalsize NUB--#1\\[-3mm]
        #2\\[-9mm]CPTH--A282.0194}\\[6mm]
\end{flushright}
\vfill
{\large \bf #3}}
\author{\large #4 \\[1cm] #5}
\maketitle
\vfill
\nopagebreak
\begin{abstract}
{\noindent #6}
\end{abstract}
\vfill
\begin{flushleft}
\rule{16.1cm}{0.2mm}\\[-3mm]
$^{\star}${\small Research supported in part by\vspace{-4mm}
the National Science Foundation under grant PHY--93--06906,
in part by the EEC contracts SC1--CT92--0792 and CHRX--CT93--0340,
\vspace{-4mm} and in part by CNRS--NSF grant INT--92--16146.}\\[-3mm]
$^{\dagger}${\small Laboratoire Propre du CNRS UPR A.0014.}\\
April 1994
\end{flushleft}
\thispagestyle{empty}
\end{titlepage}}
\newcommand{\NIJ}{{\cal N}_{IJ}}
\newcommand{\N}{{\cal N}}
\newcommand{\dal}{\raisebox{0.085cm}
{\fbox{\rule{0cm}{0.07cm}\,}}}
\newcommand{\dt}{\partial_{\langle T\rangle}}
\newcommand{\dtbar}{\partial_{\langle\fracline{T}\rangle}}
\newcommand{\al}{\alpha^{\prime}}
\newcommand{\mst}{M_{\scriptscriptstyle \!S}}
\newcommand{\mpl}{M_{\scriptscriptstyle \!P}}
\newcommand{\dv}{\int{\rm d}^4x\sqrt{g}}
\newcommand{\lv}{\left\langle}
\newcommand{\rv}{\right\rangle}
\newcommand{\ph}{\varphi}
\newcommand{\sbar}{\,\fracline{\! S}}
\newcommand{\xbar}{\,\fracline{\! X}}
\newcommand{\dslash}{\not\!\partial}
\newcommand{\barz}{\,\fracline{\! Z}}
\newcommand{\zbar}{\bar{z}}
\newcommand{\abar}{\bar{A}}
\newcommand{\bbar}{\bar{B}}
\newcommand{\chibar}{\bar{\chi}}
\newcommand{\dbar}{\,\fracline{\!\partial}}
\newcommand{\tbar}{\bar{T}}
\newcommand{\ubar}{\bar{U}}
\newcommand{\Psibar}{\overline{\Psi}}
\newcommand{\ybar}{\fracline{Y}}
\newcommand{\z}{\zeta}
\newcommand{\zb}{\bar{\zeta}}
\newcommand{\phb}{\fracline{\varphi}}
\newcommand{\cm}{Commun.\ Math.\ Phys.~}
\newcommand{\pr}{Phys.\ Rev.\ D~}
\newcommand{\pl}{Phys.\ Lett.\ B~}
\newcommand{\np}{Nucl.\ Phys.\ B~}
\newcommand{\e}{{\rm e}}
\newcommand{\gsi}{\,\raisebox{-0.13cm}{$\stackrel{\textstyle
>}{\textstyle\sim}$}\,}
\newcommand{\lsi}{\,\raisebox{-0.13cm}{$\stackrel{\textstyle
<}{\textstyle\sim}$}\,}
\date{}
\firstpage{3084}{IC/94/72}
{\large\sc Effective $\mu$-Term in Superstring Theory$^\star$}
{I. Antoniadis$^{\,a}$, E. Gava$^{b,c}$,
K.S. Narain$^{ c}$ $\,$and$\,$
T.R. Taylor$^{\,d}$}
{\normalsize\sl
$^a$Centre de Physique Th\'eorique, Ecole Polytechnique,$^{\dagger}$
F-91128 Palaiseau, France\\
\normalsize\sl
$^b$Instituto Nazionale di Fisica Nucleare, sez.\ di Trieste,
Italy\\
\normalsize\sl $^c$International Centre for Theoretical Physics,
I-34100 Trieste, Italy\\
\normalsize\sl $^d$Department of Physics, Northeastern
University, Boston, MA 02115, U.S.A.
}{In four-dimensional compactifications of the heterotic superstring
theory the K\"ahler potential has a form which generically induces
a superpotential mass term for Higgs particles once supersymmetry
is broken at low energies. This ``$\mu$-term'' is analyzed in a
model-independent way at the tree level and in the one-loop approximation,
and explicit expressions are obtained for orbifold compactifications.
Additional contributions which arise in the case of supersymmetry
breaking induced by gaugino condensation are also discussed.}
\setcounter{section}{0}
\section{Introduction}
Minimal supersymmetric generalization of the standard model of
electro-weak interactions requires the existence of two Higgs doublets.
Futhermore, a mixed mass-term is necessary for these doublets
in order to avoid a massless axion. This generates a hierarchy problem
since a new mass scale $\mu$, of order of the weak scale, must be introduced
by hand. One way to avoid this problem is that $\mu$ is induced
by the breaking of {\it local\/} supersymmetry \cite{giud}.
The $\mu$-term is then generated provided that the K\"ahler
potential mixes Higgs doublets with neutral scalars which
acquire vacuum expectation values of order of the Planck scale.

Superstring theory provides a natural setting for such a mechanism since
explicit mass terms are absent for Higgs particles while the spectrum
contains neutral moduli which can provide the required mixing
in the K\"ahler potential \cite{kl}. However, it turns out that there are
also additional contributions to the induced $\mu$-term due to effective
interactions which are not described by the standard two-derivative
supergravity. They correspond to higher weight F-terms. In this work, we
present an analysis of the moduli-dependence of these new interactions. We
also discuss the moduli-dependence of the contributions to
$\mu$ that are due to the K\"ahler potential terms
quadratic in Higgs fields. We derive
explicit expressions for orbifold compactifications, up to the one-loop level.

We will restrict our discussion to the case of (2,2) compactifications
of the heterotic superstring.
In this case, the gauge group is $E_6\times E_8$ and the matter fields
transform as {\bf 27} or $\overline{{\bf 27}}$ under $E_6$ and they are in
one-to-one correspondence with the moduli: {\bf 27}'s are related to (1,1)
moduli and $\overline{{\bf 27}}$'s to (1,2) moduli.
The K\"ahler potential has the following power expansion in
the matter fields:
\begin{equation}
K ~=~ G+A^{\alpha}A^{\bar{\alpha}}Z_{\alpha\bar{\alpha}}^{(1,1)}
+B^{\nu}B^{\bar{\nu}}
Z_{\nu{\bar\nu}}^{(1,2)} +(A^{\alpha}B^{\nu}H_{\alpha\nu}+ c.c.) +\dots\, ,
\label{Kahler}
\end{equation}
where $A$ and $B$ refer to {\bf 27}'s and $\overline{{\bf 27}}$'s, respectively.
The function $G$ defines the moduli metric which is block-diagonal in (1,1) and
(1,2) moduli: $G = G^{(1,1)} + G^{(1,2)}$. The moduli metrics as well as the
matter metrics $Z^{(1,1)}$ and $Z^{(1,2)}$ have already been studied
up to the one-loop level \cite{dkl,yuka}.
In discussing general properties
we will follow the notation of \cite{dkl}, and label ${\bf 27}$
($\overline{{\bf 27}}$) fields with letters from the beginning (middle) of the
greek alphabet. Moduli are generically labelled by latin letters.

The function $H$, which depends on the moduli,
does not affect the renormalizable couplings of
the effective field theory at energies above the
supersymmetry breaking scale. At low energies, below the supersymmetry
breaking scale, a superpotential mass term
$\mu_{\alpha\nu}A^{\alpha}B^{\nu}$ appears though
in the ``observable'' matter sector \cite{giud,kl}, with the parameter
\begin{equation}
\mu_{\alpha\nu}=m_{3/2}H_{\alpha\nu}-
h^{\bar{n}}\partial_{\bar{n}}H_{\alpha\nu}+{\tilde\mu}_{\alpha\nu}\, ,
\label{mu}
\end{equation}
where $m_{3/2}$ is the gravitino mass and
$h^{n}$ is the auxiliary component of $n$-th modulus [of (1,1) or
(1,2) variety]. The additional contribution ${\tilde\mu}_{\alpha\nu}$ depends
on the interactions discussed in the next section. It also
contains an explicit superpotential term
that may be generated at the supersymmetry breaking scale if this breaking
occurs as a result of non-perturbative effects which violate
non-renormalization
theorems. After including wave function normalization factors, the observable
fermion masses read:
\begin{equation}
m_{\alpha\nu}=Z_{\alpha\bar{\alpha}}^{-1/2}
Z_{\nu\bar{\nu}}^{-1/2}\mu_{\alpha\nu}\, .
\label{mfer}
\end{equation}

The paper is organized as follows. In section 2, we describe a higher weight
F-term which, in the presence of Yukawa couplings of charged fields
with singlets, gives rise to effective interactions that mix with the standard
two-derivative couplings. We discuss the relation between these new
interactions, Yukawa couplings and ${\tilde\mu}$. In section 3,
we examine the tree level contribution to the $H$ tensor. We express both $H$
and the coupling associated with the  higher weight interaction in terms of the
singlet Yukawa couplings, generalizing special geometry to the
matter field components of the
Riemann tensor. A significant simplification occurs in the case of orbifold
compactifications where the higher weight term vanishes identically. In section
4, we determine the dependence of the function $H$ on untwisted moduli. We show
that this dependence is consistent with large-small radius duality, with (1,2)
moduli transforming non-trivially under $SL(2,Z)$ duality transformations of
(1,1) moduli, and vice versa. In sections 5 and 6 we continue our discussion of
orbifold models. In section 5, we compute the one loop contribution to the
$H$-function. In section 6, we discuss the matter-dependent part
of the one loop threshold correction to gauge couplings. They are
phenomenologically interesting because they also induce fermion masses,
through gaugino condensation. This non-perturbative mass generation
is discussed in section 7 where explicit expressions are derived
for higgsino masses in orbifold models.

\section{A higher weight F-term}

The most convenient way of computing the $H$-function at the tree level
is to consider the four-scalar
scattering amplitude ${\cal A}(A^{\alpha}, B^{\nu},
M^{\bar m}, M^{\bar n})$ involving {\bf 27}, $\overline{{\bf 27}}$ and
two anti-moduli. In the leading, quadratic order in momenta, there
exist in general two contributions to this process.
The first, coming from the
standard D-term interactions in the effective Lagrangian, gives
the Riemann tensor of K\"ahler geometry, $R_{\alpha{\bar m}\nu{\bar n}} =
\nabla_{\bar m}\partial_{\bar n}H_{\alpha\nu}$.
The second contribution exists only in the presence of Yukawa
couplings involving $A$, $B$ and a gauge singlet $s$, and it is not described
by the standard supergravity Lagrangian.
It involves a two-derivative coupling of two anti-moduli to the auxiliary
component of $\bar s$. The auxiliary field propagates between
this coupling and the Yukawa coupling to $A$ and $B$, producing an
effective interaction which contributes to
${\cal A}(A^{\alpha}, B^{\nu},M^{\bar m}, M^{\bar n})$.
We begin this section by giving a superfield description of a higher
weight F-term which describes such higher-derivative
couplings of auxiliary fields and scalars.

The procedure that we adopt here in order
to construct locally supersymmetric F-terms
which accomodate such two-derivative mixings of auxiliary fields with
scalars and in general, interactions involving more than
two spacetime derivatives, is to construct these terms in superconformal
supergravity, and to impose subsequently the standard
$N{=}1$ Poincar\'e gauge fixing constraints \cite{kugo}. In general, multiplets
in superconformal theory are characterized by the conformal (Weyl)
and chiral weights which specify the properties under the dilatations
and chiral $U(1)$ transformations.
All (standard) chiral superfields carry
weights (0,0), except for the chiral compensator $\Sigma$ with weights
(1,1).\footnote{The first number in a bracket refers
to the Weyl weight while the second number to the chiral weight.}
An invariant action can be constructed from
the F-component of a chiral superfield with weights (3,3).
The standard superpotential $W$ is a function of (0,0) fields only, and
the corresponding lagrangian density is obtained from the
F-component of $\Sigma^3W$; in this case, the (3,3) weights are supplied
entirely by the chiral compensator.

Higher weight, generalized superpotential interactions can
be constructed by including superfields which are chiral projections
of complex vector superfields. The superconformal chiral projection
$\Pi$, which is a generalization of the $\bar{D}^2$ operator of
rigid supersymmetry,
can be defined for vector superfields $V$ with weights (2,0) only;
a chiral superfield $\Pi (V)$ has weights (3,3)
\cite{kugo}. These higher weight chiral
superfields can be used to construct F-term action densities.\footnote{
In fact, the standard two-derivative kinetic energy D-term can be written as the
F-component of $\Pi(\Sigma\overline{\Sigma}e^{-K/3})$.}
The interaction term that is relevant for the computation of the function
$H$ can be written as:
\begin{equation}
F ~=~ \left. \Sigma^{-3}\,\Pi_1\,\Pi_2\right|_{\makebox{\footnotesize
 F-component}}~, \label{F}
\end{equation}
where
\begin{equation}
\Pi_{n}\equiv\Pi(\Sigma\overline{\Sigma}e^{-K/3}f^{n})\ ,
\hskip 7mm n=1,2,
\end{equation}
with complex functions $f^1$ and $f^2$ depending on weights (0,0)
(chiral and anti-chiral) superfields only. These functions can
carry some additional indices which are assumed to be contracted in
eq.(\ref{F}).

The gauge fixing that leads to $N{=}1$ Poincar\'e supergravity is imposed
by constraining the scalar and fermionic components of the chiral compensator.
A sigma-model type constraint which is convenient for discussing
the string-loop expansion of the effective action in powers
of the string coupling $g^2$, with the coefficient
of the Ricci scalar normalized to $1/g^2$,  corresponds to
\begin{equation}
\left.\Sigma\overline{\Sigma}e^{-K/3}\right|_{\makebox{\footnotesize
scalar component}}=\frac{1}{g^2}\ .
\end{equation}
{}From the known dilaton-dependence of the K\"ahler potential it follows
then that the scalar component of $\Sigma^3\sim 1/g^2$. The F-term
(\ref{F}) is of the same order $1/g^2$ as all tree-level
interactions.\footnote{In the following, our mass units are chosen in such a
way that the coefficient of the space-time Ricci scalar becomes 1/2.}
Another property of the interaction (\ref{F}), which is useful for discussing
target space modular invariance, is that it is
invariant under K\"ahler transformations $K\rightarrow K+\ph+\bar{\ph}$,
$\Sigma\rightarrow e^{\ph /3}\Sigma$, provided that $f^1$ and $f^2$ transform
in a holomorphic way,  and
$f^1f^2\rightarrow e^{\ph}f^1f^2$.

After imposing appropriate constraints on the chiral compensator,
the components of $\Pi(\Sigma\overline{\Sigma}e^{-K/3}f)$
can be written as
\begin{eqnarray}
z_{\scriptscriptstyle\Pi} &=& \chibar^{\bar\imath}\chibar^{\bar\jmath}
\nabla_{\bar\imath}f_{\bar\jmath}
- h^{\bar\imath}f_{\bar\imath}+\dots\label{zcomp}\\
\chi_{\scriptscriptstyle\Pi} &=& f_{{\bar\imath}\bar\jmath}\dslash
z^{\bar\imath}\chibar^{\bar\jmath}+f_{\bar\imath}\dslash\chibar^{\bar\imath}
+\dots\label{chicomp}\\
h_{\scriptscriptstyle\Pi} &=& -f_{{\bar\imath}\bar\jmath}\partial_m
z^{\bar\imath}\partial_m z^{\bar\jmath}-f_{\bar\imath}\partial^2 z^{\bar\imath}
+\dots,\label{fcomp}
\end{eqnarray}
where $z^i$, $\chi^i$, $h^i$ are the scalar, fermionic and auxiliary
components, respectively, of chiral superfields ($z^{\bar\imath}$,
$\chibar^{\bar\imath}$ and $h^{\bar\imath}$ are their complex conjugates).
The subscripts on the function $f$ denote ordinary partial derivatives,
whereas the reparametrization covariant derivative $\nabla$ acts as
$\nabla_{\bar\imath}f_{\bar\jmath}=
f_{\bar\imath\bar\jmath}-\Gamma^{\bar{k}}_{\bar\imath\bar\jmath}f_{\bar{k}}$,
with the reparametrization connection $\Gamma^k_{ij}=
K^{-1\,k \bar{l}}K_{ij\bar{l}}$.
Eqs.(\ref{zcomp}-\ref{fcomp}), in which we neglected terms that are irrelevant
for the following computations, allow expressing the higher weight
F-term (\ref{F}) in terms of component fields:
\begin{eqnarray}
F&=& e^{-K/2}(\chibar^{\bar\imath}\chibar^{\bar\jmath}\nabla_{\bar\imath}
f_{\bar\jmath}^{(1}
- h^{\bar\imath}f_{\bar\imath}^{(1})\,(f_{{\bar k}\bar l}^{2)}\partial_m
z^{\bar k}\partial_m z^{\bar l}+f_{\bar k}^{2)}\partial^2 z^{\bar k})
\nonumber\\
&& + ~e^{-K/2}(f_{{\bar\imath}\bar\jmath}^{(1}\dslash
z^{\bar\imath}\chibar^{\bar\jmath}+
f_{\bar\imath}^{(1}\dslash\chibar^{\bar\imath})\,(f_{{\bar k}\bar l}^{2)}
\dslash z^{\bar k}\chibar^{\bar l}+f_{\bar k}^{2)}\dslash\chibar^{\bar k})
+\dots,\label{icomp}\end{eqnarray}
where a round bracket on the superscript denotes symmetrization
(square brackets will be used later to denote antisymmetrization).

{}From the form of the F-term in eq.(\ref{icomp}), it is clear that the
interference between these new couplings and the
$H$-dependent part of the K\"ahler potential occurs in all amplitudes involving
a {\bf 27}-$\overline{{\bf 27}}$ pair of scalars. In fact, the standard
two-derivative
$K_{AB{\bar z}}=\partial_{\bar z} H_{AB}$ vertex interferes with the
Yukawa coupling $W_{ABs}$ due to the mixing between $h_{\bar s}$ and
$\partial^2{\bar z}$. Similar interference occurs in the supersymmetric partner
of this vertex, involving two matter fermions and one auxiliary field.
The corresponding diagrams are shown on Fig.1. The diagram 1a generates the
standard contribution to $\mu$ (the second term of eq.(\ref{mu})), while 1b
gives rise to the additional contribution:
\begin{equation}
{\tilde\mu}_{AB}=-h^{\bar n}W_{ABs}G^{s{\bar s}}
f_{\bar s}^{(1}f_{\bar n}^{2)}\ ,
\label{mutilde}
\end{equation}
where $G^{s{\bar s}}$ is the inverse metric of singlets. Note that since $A$
and $B$ are massless throughout the whole moduli space, there always exists a
field basis in which there is no mixing between moduli and singlets in the
K\"ahler metric.

In the case of (2,2) models, the F-term is related to the Yukawa
couplings with singlet fields. First of all, using $N{=}(2,2)$
world-sheet supersymmetry one can show that all relevant amplitudes
originating from the F-term (\ref{F}) vanish unless they involve at least one
singlet (other than (1,1) or (1,2) moduli). Next, we consider the amplitude
${\cal A}(\chibar^{\bar\imath},\chibar^{\bar\jmath},M^{\bar{k}},\bar{s})$,
involving two anti-modulinos, one anti-modulus and one anti-singlet.
Collecting all the relevant terms in (\ref{icomp}) and using
mass-shell conditions one can show that:\footnote{Our conventions follow
ref.\cite{dkl}; in particular, the kinematic variables are defined as
$s=-(p_1+p_2)^2$,
$t=-(p_1+p_4)^2$ and $u=-(p_1+p_3)^2$, where $p_1$, $p_2$, $p_3$ and $p_4$ are
the external momenta in the order appearing in ${\cal A}$.}
\begin{equation}
{\cal A}(\chibar^{\bar\imath},\chibar^{\bar\jmath},M^{\bar{k}},\bar{s})=
{1\over 2} e^{-G/2} \{ s \nabla_{[\bar{k}}
T_{{\bar\jmath}]{\bar\imath}{\bar{s}}} + t \nabla_{[\bar\imath}
T_{{\bar\jmath}]{\bar{k}}{\bar{s}}}\}\ ,
\label{Pi24}
\end{equation}
where $T_{{\bar\imath}{\bar\jmath}{\bar{s}}}=
(\nabla_{\bar\imath}f^{(1}_{\bar\jmath}) f^{2)}_{\bar{s}}$. In the expression
(\ref{Pi24}) the reparametrization covariant derivatives appear as a result of
combining reducible and irreducible vertices coming from (\ref{icomp}) and
kinetic energy interactions.

We can now use the supercurrent of the right-moving (bosonic) sector to relate
this amplitude to that with two of the moduli being replaced by charged matter.
Note that both the world-sheet supercurrents of the internal $N{=}2$
superconformal theory have first order poles in their operator product
expansion with the singlet vertex operator. We can follow the method of
\cite{dkl} to show firstly that when all the three moduli are of $(1,1)$ or
$(1,2)$ variety this amplitude vanishes. Specializing then to the case of
$\bar\imath$ and
$\bar{k}$ refering to
$(1,1)$ and
$\bar\jmath$ to $(1,2)$ anti-moduli respectively, one obtains the following
relation:
\begin{eqnarray}
{\cal A}(\chibar^{\bar\imath},\chibar^{\bar\jmath},M^{\bar{k}},\bar{s}) &=&
\frac{\alpha'}{4}\,[s\,{\cal U}_{\bar\imath}^{\bar \alpha}\,
{\cal U}_{{\bar\jmath}}^{\bar\nu}
{\cal A}(\chibar^{\bar{\alpha}},\chibar^{\bar{\nu}},M^{\bar{k}},\bar{s})
+ t\,{\cal U}_{{\bar k}}^{\bar \beta}\,
{\cal U}_{{\bar\jmath}}^{\bar\nu}
{\cal A}(\chibar^{\bar\imath},\chibar^{\bar{\nu}},A^{\bar{\beta}},\bar{s})]
\nonumber\\
&=& \frac{1}{2} e^{-G/2}\,[ s\,{\cal U}_{\bar\imath}^{\bar \alpha}\,
{\cal U}_{{\bar\jmath}}^{\bar\nu} \nabla_{\bar k}(e^G \overline{W}_{\bar{\alpha}
\bar{\nu} \bar{s}}) + t\,{\cal U}_{\bar{k}}^{\bar \beta}\,
{\cal U}_{{\bar\jmath}}^{\bar\nu} \nabla_{\bar\imath}
(e^G \overline{W}_{\bar{\beta}
\bar{\nu} \bar{s}})]\ ,
\label{Piyuk}
\end{eqnarray}
where the Regge slope $\alpha'=2/g^2$ in our mass units.
The ${\cal U}$ matrices transform matter indices to the
corresponding moduli indices \cite{dkl}. In a natural basis for the matter
fields, they are given by ${\cal U}^{{\bar b}}_{\bar\beta} =
\delta^{{\bar
b}}_{\bar\beta} \exp{1\over 6}(G^{(1,1)}-G^{(1,2)})$ and ${\cal
U}^{{\bar\jmath}}_{\bar\mu} = \delta^{{\bar\jmath}}_{\bar\mu}
\exp{1\over 6}(G^{(1,2)}-G^{(1,1)})$. In amplitudes involving
${\bf 27}$-${\overline{{\bf 27}}}$ pairs, like in (\ref{Piyuk}),
the exponential factors cancel
and the moduli indices can be identified with matter indices.

Comparing (\ref{Pi24}) and (\ref{Piyuk}) we obtain:
\begin{equation}
\nabla_{[\bar{k}} T_{{\bar\jmath}]{\bar\imath}{\bar{s}}} =
\nabla_{[\bar{k}}(e^G \overline{W}_{{\bar\jmath}] {\bar\imath} \bar{s}})\ ,
\label{rel}
\end{equation}
where we used the fact that $\overline{W}_{{\bar\imath}{\bar{k}}\bar{s}}=0$
since both ${\bar\imath}$ and ${\bar{k}}$ correspond to $\overline{{\bf 27}}$'s.
In this way, we related the higher weight F-term (\ref{F}) to standard Yukawa
couplings involving gauge singlet fields.

In the case of orbifold compactifications, the tensor
$T_{{\bar\imath}{\bar\jmath} {\bar{s}}}$  vanishes for untwisted moduli
${\bar\imath}$ and ${\bar\jmath}$ since the singlet carries a non-zero charge
under the enhanced gauge symmetry. However, this argument does not
apply in the presence of blowing-up modes (twisted moduli)
which transform non-trivially under enhanced symmetries. Indeed we found
explicit examples in the case of $Z_6$ orbifold
in which the amplitude (\ref{Pi24}) is non-zero when all fields
belong to twisted sectors.

\section{Function $H$ at the tree level}

As already mentioned, the most
convenient way of computing $H$ is to extract it from the
scattering amplitude ${\cal A}(A^{\alpha}, B^{\nu},
 M^{\bar m}, M^{\bar n})$
involving two matter scalar fields and two anti-moduli.
There are then two possible contributions to this process
at second order in momenta: the first, coming from the
standard D-term interactions in the effective Lagrangian, gives
the Riemann tensor of K\"ahler geometry. Besides this,
there is also a contribution coming from the higher weight
interactions (\ref{F}, \ref{icomp}). The corresponding diagrams are shown
on Fig.2, and the result is:
\begin{equation}
{\cal A}(A^{\alpha}, B^{\nu}, M^{\bar m}, M^{\bar n})=
is(R_{\alpha{\bar m}\nu{\bar n}}\,-\,W_{\alpha\nu s}
G^{s\bar s}T_{{\bar m}{\bar n}{\bar s}})\ .
\label{ampl}
\end{equation}
The mixed Riemann tensor $R_{\alpha{\bar m}\nu{\bar n}}$
is directly related to the $H$-tensor. Indeed
from (\ref{Kahler})
it follows that, up to the quadratic order in matter fields,
\begin{equation}
R_{\alpha{\bar m}\nu{\bar n}} = \nabla_{\bar m}\partial_{\bar n}
H_{\alpha\nu}=
\partial_{\bar m}\partial_{\bar n}H_{\alpha\nu}
-\Gamma^{\bar{k}}_{\bar{m}\bar{n}}\partial_{\bar k}H_{\alpha\nu}\, ,
\label{rh}
\end{equation}

In the case of (2,2) compactifications, following again the approach of
ref.\cite{dkl}, one can use $N{=}2$ world-sheet supersymmetry of the
right-moving
sector to show that a non-vanishing amplitude (\ref{ampl}) must
necessarily involve two antimoduli of different (1,1) and (1,2) varieties.
Furthermore, such amplitudes can be related to amplitudes involving four matter
fields:
\begin{equation}
{\cal A}(A_{a}^{\alpha}, B_{a}^{\nu},
M^{{\bar\beta}},
M^{{\bar\mu}}) =\frac{\alpha'}{4}\, \{ s{\cal A}(A_{a}^{\alpha},
B_{a}^{\nu}, A_{b}^{{\bar\beta}}, B_{b}^{{\bar\mu}}) - t{\cal
A}(A_{a}^{\alpha},  B_{b}^{\nu}, A_{b}^{{\bar\beta}},
B_{a}^{{\bar\mu}})\}\ ,
\label{wi}
\end{equation}
where $a\neq b$ are arbitrary gauge indices.

Using the results of ref.\cite{dkl} for the four-point matter
amplitudes, one obtains:
\begin{equation}
\partial_{\bar\beta}\partial_{\bar\mu}H_{\alpha\nu} =
R_{\alpha{\bar\beta}\nu{\bar\mu}} = G_{\alpha{\bar\beta}}
G_{\nu{\bar\mu}} - W_{\alpha\nu s} G^{s{\bar s}}(e^G {\overline
W}_{{\bar\beta}{\bar\mu}{\bar s}} -T_{{\bar\beta}{\bar\mu}\bar s})\ .
\label{tree}
\end{equation}
In (\ref{tree}), we replaced the reparametrization derivative
$\nabla$ (see (\ref{rh})) by an ordinary partial derivative, since
there are no connection terms which mix (1,1) and (1,2) moduli. Equation
(\ref{tree}) generalizes special geometry to matter field coordinates of the
Riemann tensor. One can check that the Riemann tensor (\ref{tree})
obeys the Bianchi identity,
$\nabla_{\bar l}R_{\alpha{\bar m}\nu{\bar n}} =
\nabla_{\bar m}R_{\alpha{\bar l}\nu{\bar n}}$,
as a consequence of the relation (\ref{rel}) between the $T$-tensor and
the Yukawa couplings.

At this point, we have two equations, (\ref{rel}) and (\ref{tree}), which in
principle should allow expressing both tensors $H$ and $T$ in terms of the
Yukawa couplings of {\bf 27} and $\overline{{\bf 27}}$ with singlets.
We first note that
physical amplitudes like (\ref{ampl}) remain invariant under the
transformation:
\begin{equation}
T_{{\bar m}{\bar n}{\bar s}}\rightarrow T_{{\bar m}{\bar n}{\bar s}}
+\nabla_{\bar m}\nabla_{\bar n}\Lambda_{\bar s} \hskip 1cm
H_{\alpha\nu}\rightarrow H_{\alpha\nu} +W_{\alpha\nu s}G^{s{\bar s}}
\Lambda_{\bar s} \ ,
\label{inv}
\end{equation}
where $\Lambda_{\bar s}$ is an arbitrary (vector) function. The symmetry
transformation (\ref{inv}) is valid for all compactifications.
In the case of (2,2) models, this freedom can be
used to solve equation (\ref{rel}):\footnote{This solution may not be
appropriate at enhanced symmetry points, for instance in orbifold
compactifications which will be discussed in the next section; if
$T_{{\bar m}{\bar n}{\bar s}}$ is charged under the enhanced symmetry, then it
should vanish by making an appropriate choice of $\Lambda_s$.}
\begin{equation}
T_{{\bar m}{\bar n}{\bar s}}= e^G {\overline W}_{{\bar m}{\bar n}{\bar s}}\ .
\label{Tsol}
\end{equation}
Moreover, equation (\ref{tree}) provides a differential equation for $H$:
\begin{equation}
\partial_{\bar\beta}\partial_{\bar\mu}H_{\alpha\nu}= G_{\alpha{\bar\beta}}
G_{\nu{\bar\mu}}\ .
\label{Hsol}
\end{equation}

Although the transformation (\ref{inv}) is manifestly a symmetry of the
on-shell physical amplitudes, it is not obvious how this could be implemented
off-shell. This is a general problem of ambiguities in constructing the
off-shell action from the data of on-shell amplitudes. In particular, the mass
formula (\ref{mu}) with $\tilde\mu$ given in (\ref{mutilde}) is not invariant
under the transformation (\ref{inv}), as it involves couplings with auxiliary
fields. However, Yukawa couplings of the Higgs fields with singlets would give
additional contribution to $\tilde\mu$, and a non-vanishing vacuum expectation
value can be in principle generated for a singlet at the supersymmetry breaking
scale, producing a direct higgsino mass. The correct mass formula would
therefore involve the complete minimization of the scalar potential, and we
believe that the above ambiguity should disappear at the true classical
solution.

In the case of compactifications which give rise to the particle content
of the minimal supersymmetric standard model at low energies, the singlets are
either superheavy or have no Yukawa couplings with Higgs particles. In both
cases, the additional contribution (\ref{mutilde}) vanishes and the induced
$\mu$-term (\ref{mu}) depends entirely on the function $H$ and
eventual non-perturbative superpotential. Furthermore, equation (\ref{Hsol})
can be integrated to determine the function $H$.

\section{Tree level $\mu$-term in orbifold compactifications}

As an example, we determine the dependence of the function $H$ on untwisted
moduli in orbifold compactifications. In this case, equation (\ref{tree}) could
be in principle modified due to exchange of gauge bosons of the enhanced
symmetry group in amplitudes involving four matter fields. However, using the
results of ref.\cite{dkl}, one can show that these additional contributions
cancel in the combination of four matter field
amplitudes appearing on the r.h.s.\ of
eq.(\ref{wi}). Therefore, relation (\ref{tree}) remains valid in the case of
orbifolds, as well. Furthermore, in the case of untwisted moduli, the Yukawa
couplings appearing in eq.(\ref{tree}) are constants (moduli independent)
while as already mentioned, the higher weight F-term is absent. Thus, the
computation becomes very simple.

Consider first untwisted matter fields
which can be parametrized as matrices $A^{\alpha_{L}
\alpha_{R}}$ and
$B^{\beta_L{\bar\beta}_R}$ where $L$- and $R$-indices refer to the three
left- and right-moving (complex) internal coordinates. In fact, $L$-indices
label the three complex orbifold planes whereas $R$-ones
correspond to gauge indices of the enhanced symmetry group. A bar on a
$R$-index is associated with $\overline{{\bf 27}}$'s and (1,2) moduli, while a
bar on a $L$-index will represent ordinary complex conjugation. The vertex
operators (in the 0--ghost picture) for matter fields are:
\begin{eqnarray}
A^{\alpha_L\alpha_R} &:& (\partial X^{\alpha_L} + ip\cdot\psi \Psi_L^{\alpha_L})
\Psi_R^{\alpha_R}\lambda e^{ip\cdot x}\nonumber\\
B^{\beta_L{\bar\beta}_R} &:& (\partial X^{\beta_L} +
ip\cdot\psi \Psi_L^{\beta_L})
\Psi_R^{*{\bar\beta}_R}\lambda e^{ip\cdot x}\ ,
\label{matterv}
\end{eqnarray}
where $x$ and $X$ represent the bosonic space-time and internal coordinates,
respectively, while $\psi$ and $\Psi_L$ are their left-moving fermionic
superpartners. $\Psi_R$ are the right-moving fermionic coordinates and
$\lambda$ are the $E_6$ word-sheet fermions. The corresponding (1,1) and (1,2)
moduli vertices are obtained from (\ref{matterv}) by replacing
$\Psi_R^{\alpha_R}\lambda$ and $\Psi_R^{*{\bar\beta}_R}\lambda$ by
${\bar\partial} X^{\alpha_R}$ and ${\bar\partial} X^{*{\bar\beta}_R}$,
respectively.

The relevant amplitude (\ref{ampl}), involving the above matter fields together
with one (1,1) anti-modulus $\tbar^{{\bar\gamma}_L\gamma_R}$ and one (1,2)
anti-modulus $\ubar^{{\bar\delta}_L{\bar\delta}_R}$, can be directly computed.
Transforming the two matter vertices to the $-1$--ghost picture, one finds:
\begin{eqnarray}
{\cal A}(A^{\alpha_L\alpha_R}, B^{\beta_L{\bar\beta}_R},
\tbar^{{\bar\gamma}_L\gamma_R},
\ubar^{{\bar\delta}_L{\bar\delta}_R}) &=& is \delta^{\alpha_R{\bar\beta}_R}
\delta^{\gamma_R{\bar\delta}_R} \int \frac{d^2 z}{\pi}
|z|^{-\alpha'u/2} |1-z|^{-\alpha't/2} \nonumber\\
&&~\times
\{ \frac{1}{z} G^{(1,1)}_{\alpha_L{\bar\gamma}_L}
G^{(1,2)}_{\beta_L{\bar\delta}_L} +
\frac{1}{1-z} G^{(1,1)}_{\alpha_L{\bar\delta}_L}
G^{(1,2)}_{\beta_L{\bar\gamma}_L} \}\nonumber\\
&&\hskip -3cm  ~=~ -i \delta^{\alpha_R{\bar\beta}_R}
\delta^{\gamma_R{\bar\delta}_R}\,
(t\, G^{(1,1)}_{\alpha_L{\bar\gamma}_L} G^{(1,2)}_{\beta_L{\bar\delta}_L} +
u\, G^{(1,1)}_{\alpha_L{\bar\delta}_L} G^{(1,2)}_{\beta_L{\bar\gamma}_L}) +
{\cal O}(\alpha')\ ,
\label{treeorb}
\end{eqnarray}
where the $SL(2,C)$ invariance was used to fix the world-sheet positions of
$\tbar$, $\ubar$ and $A$ vertices at 1, 0 and $\infty$, respectively. It is a
general property of orbifold compactifications that non-diagonal moduli
matrices are in one-to-one correspondence with non-singlet representations of
non-abelian enhanced symmetry groups; furthermore $U$ and $B$ matrices are
always diagonal. It follows from (\ref{treeorb}) that the amplitude under
consideration is non zero only if all moduli and matter fields are associated
with the same plane, which implies that the matter fields are singlets with
respect to non-abelian enhanced symmetries. Evaluating (\ref{treeorb}) and
using relations (\ref{ampl}) and (\ref{rh}) one obtains:
\begin{equation}
\partial_{\tbar} \partial_{\ubar} H_{AB} = G_{T\tbar} G_{U\ubar}\ ,
\label{hunt}
\end{equation}
where we used the fact that there is no connection that mixes (1,1) and (1,2)
moduli.

This result agrees with eq.(\ref{tree}), since the Yukawa couplings vanish for
untwisted fields from the same plane, and the tensor $T$ is zero. Furthermore,
one can easily check that the zero result in the case when some fields are
associated with different planes is also consistent with eq.(\ref{tree}): the
contribution of Yukawa couplings cancels against the contribution of the
metrics term. Using the known expressions for the moduli metrics,
$G_{T\tbar}=1/(T+\tbar)^2$, $G_{U\ubar}=1/(U+\ubar)^2$, and the tree-level
Peccei-Quinn
continuous symmetry for the pseudoscalar components of $T$ and
$U$, valid for amplitudes involving untwisted fields,
one can integrate eq.(\ref{hunt}) to obtain
\begin{equation}
H_{AB}=\frac{1}{(T+\tbar)(U+\ubar)}\ .
\label{hab}
\end{equation}
Once supersymmetry is broken, the induced $\mu$-term of equation
(\ref{mu}) yields a mass (\ref{mfer}) for the higgsinos:
\begin{equation}
m = m_{3/2} + (T+\tbar)h_T + (U+\ubar)h_U +
(T+\tbar)(U+\ubar){\tilde\mu}_{AB}\ ,
\label{mh}
\end{equation}
where we used the tree-level expressions $Z_{\abar A}=Z_{\bbar B}=[(T+\tbar)
(U+\ubar)]^{-1}$. In this case ${\tilde\mu}_{AB}=e^{G/2}W_{AB}$, where $W_{AB}$
represents an explicit superpotential-mass induced by non-perturbative effects.
The gravitino mass $m_{3/2}=e^{G/2}W$, where $W$ is the superpotential.

There exists yet another way of deriving eq.(\ref{hab}).
The effective action describing untwisted moduli and matter fields
can be obtained by the orbifold truncation of $N{=}4$ supergravity
Lagrangian along the lines of ref.\cite{fp}.
A plane containing (1,2) modulus $U$ and (1,1) modulus $T$
contributes to the K\"ahler potential as
\begin{equation}
K=-\ln [(T+\tbar)(U+\ubar)\,-\,(A+\bbar)(\abar+B)]\ ,
\label{logy}
\end{equation}
which yields the result (\ref{hab}) for the function $H$, when expanded to the
quadratic order in the matter fields.

The K\"ahler potential of eq.(\ref{logy}) is consistent with the
$SL(2,Z)$ large-small compactification radius duality:
\begin{equation}
T \rightarrow {aT-ib\over icT+d}
\label{dual}
\end{equation}
where, as usual, $ad-bc=1$. The untwisted matter fields transform under
this transformation as $SL(2,Z)$ modular forms of weight 1:
\begin{equation}
A\rightarrow (icT+d)^{-1}A\, , ~~~~~~~B\rightarrow (icT+d)^{-1}B\, .
\label{abdual}\end{equation}
$SL(2,Z)$ duality does indeed induce a K\"ahler symmetry transformation
provided that the associated (1,2) modulus transforms as
\begin{equation}
U\rightarrow U-{ic\over icT+d}\,AB\ .
\label{utran}
\end{equation}
Note that in the presence of matter fields $U$ is no longer inert under
large-small radius duality. Similarly, $T$ transforms under $SL(2,Z)$
$U$-duality.

Consider now the case of twisted matter fields $A$ and $B$. As we have
previously explained, the relevant Riemann tensor
$R_{A{\bar T}B{\bar U}}=\partial_{\bar T}\partial_{\bar U}H_{AB}$ is still
subject to eq.(\ref{tree}), with the tensor $T_{{\bar\beta}{\bar\mu}\bar s}=0$.
Since $A$ and $B$ are twisted while ${\bar T}$ and
${\bar U}$ are untwisted, the first term on the r.h.s.\ of eq.(\ref{tree})
involving the moduli metrics $G_{A{\bar T}}G_{B{\bar U}}$ vanishes, and the
Riemann tensor is given only by the second term involving the Yukawa couplings
with $E_6$ singlets:
\begin{equation}
\partial_{\bar T}\partial_{\bar U}H_{AB}=
-e^G W_{ABs}G^{s{\bar s}}\overline{W}_{{\bar T}{\bar U}{\bar s}}\, .
\label{ht}\end{equation}
Since the moduli $T$ and $U$ are untwisted, the singlets $s$ appearing in
the above sum should also be untwisted fields, otherwise the Yukawa couplings
$\overline{W}_{{\bar T}{\bar U}{\bar s}}$ are zero.
On the other hand, these singlets
should be neutral under all gauge symmetries. The reason is that $A$ and
$B$ must have opposite charges under the orbifold enhanced gauge symmetry
$U(1)^2$, otherwise the four-point amplitude ${\cal A}(A,B,{\bar T},{\bar U})$
would vanish by charge conservation, as the untwisted moduli $T$ and $U$ are
neutral under all gauge symmetries. However, such untwisted singlet fields
which are neutral under all enhanced gauge symmetries do not exist. It
follows that the Riemann tensor (\ref{ht}) and, consequently, $H_{AB}$ vanish
in the case of twisted matter fields, at the string tree-level.

\section{One Loop Corrections to the function $H$}

A general method for computing one-loop corrections to the K\"ahler
potential has been described in ref.\cite{yuka}. In particular, explicit
expressions for the one-loop corrections to the K\"ahler metric have been
derived in orbifold compactifications by considering the couplings of matter
and moduli fields to the antisymmetric tensor $b^{\mu\nu}$. Here, we shall
follow a similar method to compute the one-loop correction $H^{(1)}_{AB}$ to
the function $H$. In fact \cite{yuka}:
\begin{eqnarray}
{\cal A}(A^{\alpha_L\alpha_R}, B^{\beta_L{\bar\beta}_R},
\ubar^{{\bar\gamma}_L\gamma_R},
b^{\mu\nu})&=& i\epsilon^{\mu\nu\lambda\rho} (p_{1\lambda}+p_{2\lambda})
p_{3\rho} [\partial_{\ubar} H^{(1)}_{AB} +{1\over U+\ubar} G^{(1)}_{T\ubar}
+{1\over T+\tbar} G^{(1)}_{U\ubar}]\nonumber\\
& &\hspace{1cm} +~ {\cal O}(\alpha')\,.
\label{bmn}
\end{eqnarray}
Note that the last two contributions in the r.h.s.\ correspond to reducible
diagrams involving an intermediate modulus field propagating
in the $s$-channel coupled to $AB$ at the tree-level and to $b\ubar$ at
one loop.\footnote{The existence of one-loop corrections to the K\"ahler metric
$G^{(1)}_{k\bar{k}}$ of the moduli associated with $Z_2$-twisted planes has
been recently pointed out to us by V. Kaplunovsky. These arise due to
singularities associated with additional massless particles at $T=U$ which have
been overlooked in ref.\cite{yuka}.}

In order to extract $H^{(1)}_{AB}$ it is sufficient to consider only the odd
spin structure contribution to this amplitude. To perform the computation, we
use the form (\ref{matterv}) for the  matter and moduli vertex operators
together with the vertex operator for the antisymmetric tensor in the
$-1$--ghost picture, $\psi^{[\mu}{\bar\partial}x^{\nu]}e^{ip_4\cdot x}$,
accompanied by one supercurrent insertion. A simple zero mode counting
shows that, up to the
quadratic order in the external momenta, only $N{=}2$ orbifold sectors
contribute to this amplitude. In fact, $N{=}4$ sector requires 10 left-moving
fermionic zero modes, whereas in $N{=}1$ sectors only the zero momentum
part of
one of the matter fields contributes to this order, and the corresponding
left-moving boson $\partial X$ cannot be contracted.

In $N{=}2$ sectors, 2 left-moving fermions of the untwisted plane
are saturated by the zero modes, as well as 4 space-time fermionic coordinates,
and a non vanishing result is obtained only when all three fields $A$, $B$ and
$\ubar$ come from the same, untwisted plane. The result is:
\begin{eqnarray}
\partial_{\ubar}H^{(1)}_{AB}&+&{1\over U+\ubar} G^{(1)}_{T\ubar}
+{1\over T+\tbar} G^{(1)}_{U\ubar} ={1\over 8(2\pi)^4} {1\over
(T+\tbar)(U+\ubar)^2}\nonumber\\
&&\times\int {d^2\tau\over\tau_2}
\sum_{p_L,p_R} p_L{\bar p}_R q^{\frac{1}{2}|p_L|^2}\bar{q}^{\frac{1}{2}|p_R|^2}
\bar{\eta}(\bar{\tau})^{-2}
\int d^2 z\, \langle
\Psi_R (\bar{z})\bar{\Psi}_R(0) \lambda (\bar{z})\lambda (0) \rangle\nonumber\\
&=&{-1\over 8(2\pi)^5} {1\over (U+\ubar)^2} \partial_{T} \int {d^2\tau\over
\tau_2^2} Z
\bar{\eta}(\bar{\tau})^{-2}\, \int d^2 z\, \langle
\Psi_R (\bar{z})\bar{\Psi}_R(0) \lambda (\bar{z})\lambda (0) \rangle\ ,
\label{h1}
\end{eqnarray}
where $\tau = \tau_1+i\tau_2$ is the Teichm\"uller parameter of the world-sheet
torus, $q=e^{2\pi i\tau}$, and $\eta$ is the Dedekind eta function. $p_L$ and
$p_R$ are the left and right momenta, respectively, associated with the
untwisted plane:
\begin{equation}
p_L=\frac{1}{\sqrt{(T+\tbar)(U+\ubar)}}(m_{1}+i m{_2} \ubar
+i n{_1} \tbar-n_{2}\ubar\tbar)
\label{mom}
\end{equation}
with integer $n_1$, $n_2$, $m_1$ and $m_2$. $p_R$ is obtained by replacing
$\tbar$ with ${-}T$. $Z$ is the lattice partition function,
$Z=\sum_{p_L,p_R} q^{|p_L|^2/2} \bar{q}^{|p_R|^2/2}$. The second equality in
(\ref{h1}) has been derived by using the identity:
\begin{equation}
\partial_T Z={2\pi\tau_2\over{T+\tbar}} \sum_{p_L,p_R} p_L{\bar p}_R
q^{\frac{1}{2}|p_L|^2}\bar{q}^{\frac{1}{2}|p_R|^2}\ .
\label{Z}
\end{equation}

The integrals involved in (\ref{h1}) have already been evaluated in
\cite{yuka} when computing the one-loop  correction to the K\"ahler metric.
First, the $z$-integral gives:
\begin{equation}
{1\over (2\pi)^2}\frac{1}{\tau_2}\int d^2 z\, \langle
\Psi_R (\bar{z})\bar{\Psi}_R(0) \lambda (\bar{z})\lambda (0) \rangle
=-\makebox{Tr} (\Gamma_A-{1\over 4\pi\tau_2}) {\bar q}^{{\bar L}_0 -{11\over
12}}
\label{zint}
\end{equation}
where $\Gamma_A$ is the anomalous dimension operator defined in \cite{yuka},
and the trace extends over the right movers except for the space-time
coordinates and the lattice of the untwisted plane.
Finally, the integral over $\tau$ in (\ref{h1}) has also
been computed in \cite{yuka}, where it was shown to be
proportional to the wave function renormalization
factors, $Y_A$, $Y_B$:
\begin{equation}
Y_A=Y_B=-{\tilde{b}}_A\ln[|\eta(iT)\eta(iU)|^4(T+\tbar)(U+\ubar)]
+G^{(1)}+y_A\, .
\label{yu}\end{equation}
Here, $\tilde{b}_A={\hat{b}_A/ ind}$, where ${\hat{b}}_A$ is equal
to the $\beta$-function
coefficient of the gauge group that transforms $A$ and $B$ non-trivially in the
corresponding $N{=}2$ supersymmetric orbifold, and $ind$ is the index of the
little subgroup of the untwisted plane in the full orbifold group \cite{dkl2}.
Furthermore, $y_A$ is a constant which does not depend on the
moduli $T$ and $U$.
In this way, one finds that the
r.h.s.\ of (\ref{h1}) is equal $(-{\tilde{b}}_A G_2(T) +G^{(1)}_T)/(U+\ubar)^2$,
where the function $G_2(T)\equiv 2\partial_T\ln\eta (iT) + 1/(T+\tbar)$.
As a result, we obtain the following one-loop correction to the function $H$:
\begin{equation}
H^{(1)}_{AB} = {\tilde{b}}_A G_2(T) G_2(U) -{1\over U+\ubar}G^{(1)}_T
-{1\over T+\tbar}G^{(1)}_U\ .
\label{h1AB}
\end{equation}
Note that since $G^{(1)}$ is invariant under
the transformation (\ref{dual}), the above expression is consistent with
the invariance of the one loop correction to the K\"ahler potential,
$G^{(1)}+(H^{(1)}_{AB}AB+c.c.)$, under
the full set of $SL(2,Z)$ duality transformations (\ref{dual}-\ref{utran}).

The one loop higgsino mass,
or more precisely, its boundary value at the unification scale $M$,
can be obtained from eq.(\ref{mfer}):
\begin{eqnarray}
m(M) &=& (T+\tbar )(U+\ubar )[1+g^2(Y_A+Y_B)]^{-1/2} \nonumber\\
&&\hspace*{-12mm} \times\{ m_{3/2}(H_{AB}+g^2H^{(1)}_{AB})
-\sum_{\bar{n}=\tbar,\ubar}h^{\bar{n}}\partial_{\bar{n}}(H_{AB}
+g^2H^{(1)}_{AB})+e^{(G+g^2G^{(1)})/2}W_{AB}\},
\label{mloop}
\end{eqnarray}
where the tree-level and the one-loop contributions to the function $H$
are given in eqs.(\ref{hab}) and (\ref{h1AB}), respectively.


In order to examine the modular transformation properties
of the higgsino mass (\ref{mloop}),
it is convenient to expand the superpotential as:
\begin{equation}
W=W_0+W_{AB}AB+\dots
\label{wind}\end{equation}
$SL(2,Z)$ invariance of the effective action under the
transformations (\ref{dual}-\ref{utran}) requires that the superpotential
transforms as
$W\rightarrow (icT+d)^{-1} W$, hence
\begin{equation}
W_0\!\rightarrow\! (icT+d)^{-1} W_0\ ,\hskip 14mm
W_{AB}\! \rightarrow\!(icT+d)W_{AB}+ic\, \partial_UW_0\, .
\label{wabdual}\end{equation}
Using auxiliary field equations $h^{\bar{n}}=m_{3/2}
K^{\bar{n}n}(\partial_n\ln W+K_n)$ and eqs.(\ref{wabdual}), one finds
\begin{equation}
m(M)\rightarrow \left( {icT+d\over -ic\tbar+d}\right)^{1/2} m(M)\, ,
\label{mdual}\end{equation}
which shows that the physical mass transforms with an unobservable
phase factor.

\section{Matter field dependence of threshold corrections to gauge couplings}

We will consider in this section the $AB$-dependent part of one-loop
threshold corrections to gauge couplings. Although a general discussion can be
done, we will consider here for simplicity only the case of orbifolds.

It is convenient to start from the CP-odd axionic
couplings, corresponding to amplitudes of the type
${\cal A}(A,B,A_{\mu},A_{\nu})$, where $A_{\mu}$ and $A_{\nu}$ are gauge
bosons. To simplify the discussion, we consider the case of gauge
indices for the fields in a way that there no direct gauge interactions
between $A$, $B$ and the gauge bosons. This amplitude receives
contribution only from the odd spin-structure on the torus.
By taking into account all the reducible diagrams, one can
write:
\begin{equation}
{\cal A}(A^{\alpha_L\alpha_R}, B^{\beta_L{\bar\beta}_R},A_{\mu},
A_{\nu})_{odd}~=~i\epsilon_{\mu\nu\rho\sigma}p_{3}^{\rho}
p_{4}^{\sigma}(\Theta_{AB}^{(1)} -H_{AB,\bar k}G^{{\bar k}k}
\Theta^{(1)}_{k})
\label{tetaab}
\end{equation}
where the irredicible term $\Theta_{AB}^{(1)}$ is the quantity we are looking
for, and $\Theta^{(1)}_k$ is the usual one-loop axionic coupling of $k$-th
modulus which gives rise to a reducible diagram through the tree-level
K\"ahler vertex of $A$, $B$ and $k$-th anti-modulus. Note that another
possible reducible diagram, involving an intermediate singlet coupling to $AB$
via Yukawa, vanishes for orbifolds, as the singlet is charged under the
enhanced  gauge groups.

Proceeding now as in the previous section, we find that
$N{=}4$ and $N{=}1$ sectors do not contribute, while $N{=}2$
sectors contribute only
in the case in which $A$ and $B$ come from the
same untwisted plane. We then derive the following expression:
\begin{eqnarray}
{\cal A}(A^{\alpha_L\alpha_R}, B^{\beta_L{\bar\beta}_R},A_{\mu},
A_{\nu})_{odd}~&=&~{1\over 32{\pi}^3}{\epsilon_{\mu\nu\rho\sigma}p_{3}^{\rho}
p_{4}^{\sigma}\over (T+\tbar)(U+\ubar)}\int {d^2\tau\over \tau_2}
\sum_{p_L,p_R} {p_L}^{2} q^{\frac{1}{2}|p_L|^2}\bar{q}^{\frac{1}{2}|p_R|^2}
\nonumber\\
&&\hskip -1.5cm\times\bar{\eta}(\bar{\tau})^{-2} \int d^2 z\, \langle
(Q{^2}-\frac{1}{4\pi\tau_2})
\psi_R (\bar{z})\bar{\psi}_R(0) \lambda (\bar{z})\lambda (0) \rangle
\label{tetaab1}
\end{eqnarray}
where $Q$ is the gauge charge operator. The expression $(Q{^2}-1/{4\pi\tau_2})$
comes from the integration of the two-point function of the (level one)
Kac-Moody currents
\cite{agn}. The remaining $z$-integral in (\ref{tetaab1}) is the same as
that of (\ref{zint}).

To proceed further, we take a derivative with respect to $\ubar$ and make use of
the explicit expressions for $p_L$, $p_R$ in terms of $U$ and $T$ given in
eq.(\ref{mom}). This gives:
\begin{eqnarray}
(U+\ubar)^2\partial_{\ubar}\sum_{p_L,p_R}{{p_L}^{2}\over
(U+\ubar)}q^{\frac{1}{2}|p_L|^2}
\bar{q}^{\frac{1}{2}|p_R|^2}&=& -\frac{2i}{\tau_2}
\partial_{\tau}\sum_{p_L,p_R} {\tau_2}^{2}p_{L}{\bar
p}_{R}q^{\frac{1}{2}|p_L|^2}
\bar{q}^{\frac{1}{2}|p_R|^2}\nonumber\\
&=& -i\frac{T+\tbar}{\pi\tau_2}\partial_{\tau}
(\tau_{2}\partial_{T}Z)\, ,
\label{ident}
\end{eqnarray}
where in the second step we used the identity (\ref{Z}).
Using (\ref{ident}) and (\ref{zint}) we obtain:
\begin{eqnarray}
\partial_{\ubar}(\Theta_{AB}^{(1)} &-& H_{AB,\bar k}G^{{\bar k}k}
\Theta^{(1)}_{k})\nonumber\\
&=& {1\over 8\pi^2}{1\over (U+\ubar)^2}\int
{d^2\tau}(\partial_{\tau}\tau_{2}\partial_{T}Z)\bar{\eta}^{-2}\,
\makebox{Tr}(Q^2-{1\over 4\pi\tau_2}) (\Gamma_A -{1\over 4\pi\tau_2})
\nonumber\\
&=& {i\over 8(2\pi)^3}{1\over (U+\ubar)^2}\partial_{T}\int
{d^2\tau\over\tau_2}Z\bar{\eta}^{-2}\,
\makebox{Tr}[(Q^2-{1\over 4\pi\tau_2}) + (\Gamma_A -{1\over 4\pi\tau_2})]\,.
\label{delu}
\end{eqnarray}
In the second equality of the above equation, we have used the fact that the
boundary term for $\tau_{2}\rightarrow\infty$ vanishes due to the presence of
$\partial_{T}$, for generic radii.

The second term on the r.h.s.\ of (\ref{delu}) is identical to
$i[\partial_{\ubar}H^{(1)}_{AB} +{1\over U+\ubar}G^{(1)}_{T\ubar} +{1\over
T+\tbar}G^{(1)}_{U\ubar}]$, as can be seen from (\ref{h1}) and  (\ref{zint}).
The first term is proportional to $\Theta^{(1)}_T$:
\begin{equation}
\Theta^{(1)}_{T}\equiv {-i\over 8(2\pi)^3}\partial_{T}\int
{d^2\tau\over\tau_2}Z\bar{\eta}^{-2}\,
\makebox{Tr}(Q^2-{1\over 4\pi\tau_2})
~=~i[-{\tilde b}G_{2}(T)+G^{(1)}_T]\, ,
\label{thetat}
\end{equation}
where ${\tilde b}={\hat b}/ind$ with $\hat b$ being the $\beta$-function
coefficient of the gauge group associated to the gauge fields $A_{\mu}$,
$A_{\nu}$ in the corresponding $N{=}2$ supersymmetric orbifold. Combining the
two terms, one finds:
\begin{equation}
\partial_{\ubar}(\Theta_{AB}^{(1)}-H_{AB,\bar k}G^{{\bar k}k}
\Theta^{(1)}_{k}-iH_{AB}^{(1)})~=~
i\frac{\tilde b}{(U+\ubar)^{2}}G_{2}(T) +i\partial_{\ubar}
({1\over U+\ubar}G^{(1)}_T +{1\over T+\tbar}G^{(1)}_U)\,.
\label{equ}
\end{equation}

One can integrate this equation using duality invariance, with
the result:
\begin{equation}
\Theta_{AB}^{(1)}-H_{AB,\bar k}G^{{\bar k}k}
\Theta^{(1)}_{k}-iH_{AB}^{(1)}~=~-i{\tilde b} G_{2}(T)G_{2}(U)
+i({1\over U+\ubar}G^{(1)}_T +{1\over T+\tbar}G^{(1)}_U)\,.
\label{sol}
\end{equation}
Using the results of section 4 on the tree-level $H_{AB}$, the modulus $k$ in
(\ref{sol}) runs over $T$ and $U$. Substituting the expression (\ref{thetat})
for $\Theta^{(1)}_{T}$ (and similarly for $\Theta^{(1)}_{U}$ by replacing $T$
with $U$), one obtains:
\begin{eqnarray}
\Delta^{(1)}_{AB} &=& -{\tilde b}[G_{2}(T)G_{2}(U)-\frac{1}{(U+\ubar)}
G_{2}(T)-\frac{1}{(T+\tbar)}G_{2}(U)] + H^{(1)}_{AB}
\nonumber\\
&=& {\tilde b} [{1\over (T+\tbar)(U+\ubar)} -
4\partial_T\ln\eta (iT)\partial_U\ln\eta (iU)] + H^{(1)}_{AB}\, ,
\label{final}
\end{eqnarray}
where $\Delta$ is the inverse gauge coupling squared related to the axionic
couplings by $\Delta_T=-i\Theta_T$ \cite{dkl2}.
As expected, $\Delta^{(1)}_{AB}$ does not
transform covariantly under duality transformations.
In fact, as shown in section 4, under duality transformation
$T\rightarrow 1/T$, $U$ transforms inhomogeneously
by an additive term proportional to $AB$. It is then the
combination $\Delta^{\makebox{\scriptsize 1-loop}}
\equiv\Delta^{(1)}+(\Delta^{(1)}_{AB}AB+c.c.)+\ldots$ which
is duality invariant as can be explicitly verified.

{}From the second equality in (\ref{final}), note that the one loop correction
to the $H$ tensor, $H^{(1)}_{AB}$, appears as a universal (gauge group
independent) term of threshold corrections, expanded to first order in matter
fields. This is similar to the case of the leading order term of threshold
corrections, where the universal piece (Green-Schwarz term \cite{anom})
is again given
by the one-loop correction to the K\"ahler moduli metric \cite{yuka}.

Differentiation of (\ref{final}) with respect to $\tbar$ gives:
\begin{equation}
\partial_{\tbar}\Delta^{(1)}_{AB}={\tilde b}\partial_{\tbar}H_{AB}
+\partial_{\tbar} H^{(1)}_{AB}\,.
\label{recrel}
\end{equation}
In fact, by following the method of ref.\cite{agn} one can show that this
equation is valid for arbitrary (2,0) compactifications, when there are no
massless matter fields transforming non-trivially under the gauge group.
In this case ${\tilde b}=-c_a$, where $c_a$ is the quadratic Casimir
of the adjoint representation.
It has been shown previously \cite{yuka} that, to the leading order in matter
fields, the one-loop threshold corrections satisfy
\begin{equation}
\partial_{\bar\imath} \partial_{j} \Delta^{\makebox{\scriptsize 1-loop}}
 ={\tilde b}K^{(0)}_{{\bar\imath}j}+K^{(1)}_{{\bar\imath}j}\ ,
\label{thresh}
\end{equation}
where the indices ${\bar\imath}$, $j$ represent fields which are
neutral under the gauge group associated with
$\Delta^{\makebox{\scriptsize 1-loop}}$.
Eq.(\ref{recrel}) implies that the validity of (\ref{thresh})
extends to higher orders in matter fields.
This result could be anticipated on purely field-theoretical grounds
\cite{anom,ft}: the first term on the r.h.s.\ of
(\ref{thresh}) is related to anomalous graphs involving the coupling of the
K\"ahler current to gauginos, whereas the second term
is due to the form of the Green-Schwarz term
which contributes to
both K\"ahler potential and gauge couplings \cite{yuka}.

\section{Fermion masses induced by gaugino condensation}

In this section we discuss some phenomenological implications
of matter field dependent threshold corrections discussed in the previous
section, for the gaugino condensation mechanism of supersymmetry breaking.
Here again, we give an explicit discussion for orbi\-folds,
however some general features will be shared by all compactifications.
The analytic part of matter field dependent threshold
corrections can be always expressed as the F-term of $AB\,{\cal W}^2$ times some
analytic function of the moduli; here ${\cal W}$ is the usual gauge
field-strength
superfield. It is clear that gaugino condensation, that is a non-vanishing
vacuum expectation value of the scalar component of ${\cal W}^2$,
induces a direct mass term for $A$ and $B$.

Non-perturbative formation of gaugino condensates can be understood
in terms of the effective theory of a gauge-singlet
composite supermultiplet of gauginos and gauge bosons, coupled to
``elementary'' fields like the dilaton, moduli and matter fields \cite{fmtv}.
This composite field can be integrated out, giving rise to an effective
superpotential for the remaining fields \cite{lt}.
The resultant effective action is highly constrained by the large-small
compactification radius symmetry, which can be used very efficiently
to determine the moduli-dependence of non-perturbative superpotential
\cite{filq}.

The common feature of effective superpotentials induced by gaugino
condensation is the presence of the non-perturbative factor
$\exp (3S/2b_0)$, where $S$ is the dilaton super\-field
and $b_0$ is the one-loop
$\beta$-function coefficient of the ``hidden'' gauge
group which, in order to make our discussion explicit, we assume to be $E_8$;
in this case, $b_0=3{\tilde b}=-90$.
In order for the effective action to be invariant under the $SL(2,Z)$
transformations (\ref{dual}-\ref{utran}), the superpotential must transform as
in eqs.(\ref{wabdual}).
Although the dilaton does not transform under target space duality
transformations at the tree level, at the one loop level its
transformation properties depend on the form of the Green-Schwarz term.
As pointed out in section 5, the one-loop correction to
the K\"ahler potential and the corresponding Green-Schwarz term
are invariant under the transformations (\ref{dual}-\ref{utran}),
up to quadratic order in matter fields.
This implies that, up to this order, the dilaton
remains inert under the $SL(2,Z)$ duality transformations.

The non-perturbative superpotential that depends on the dilaton
in the right way, and satisfies the symmetry requirements (\ref{wabdual}),
is
\begin{equation}
W ~=~ e^{S/2{\tilde b}}\,\eta^{-2}(iT)\eta^{-2}(iU)\;[1-4AB\,
\partial_T\ln\eta(iT)
\partial_U\ln\eta(iU)]\;\widetilde{W} ~+\;{\cal O}[(AB)^2],\label{wnonp}
\end{equation}
where $\widetilde{W}$ may depend on the moduli of the two
other planes.\footnote{
To be consistent, this superpotential should be used with the one-loop
corrected K\"ahler potential $K=-\ln(S+\bar{S}-2G^{(1)}-
2H^{(1)}_{AB}AB+\dots)+\dots$}
Note that if the modular anomaly \cite{anom,ft}
associated with a given plane
is fully cancelled by the universal Green-Schwarz term (i.e.\
in the absence of ``analytic'' threshold corrections),
$\widetilde W$ does not depend on the corresponding moduli
since the full modular weight is provided by the
$e^{S/2{\tilde b}}$ factor in the superpotential.
The second term inside the bracket in eq.(\ref{wnonp}),
which from the point of view of
non-perturbative dynamics originates from matter field-dependent
threshold corrections (\ref{final}), gives rise to a direct mass for Higgs
particles.
Using auxiliary field equations
and eq.(\ref{mh}), we obtain $h_T=-m_{3/2}G_2(T)+{\cal O}(g^2)$,
$h_U=-m_{3/2}G_2(U)+{\cal O}(g^2)$, and the higgsino mass
\begin{equation}
m=-m_{3/2} (T+\tbar)(U+\ubar)G_2(T)G_2(U) ~+~{\cal O}(g^2)\ ,
\label{mnonp}
\end{equation}
where the gravitino mass $m_{3/2}=e^{G/2}W$.

Non-perturbative generation of higgsino masses is an inevitable
consequence of the transformation rules of $SL(2,Z)$ duality symmetry
which involve matter-field dependent, inhomogenous terms
as in eq.(\ref{utran}).
In order to obtain concrete values of higgsino masses, the vacuum expectation
values of the moduli should be determined
by minimizing the effective potential. Unfortunately,
the superpotential (\ref{wnonp}) suffers from the usual dilaton instability,
therefore a complete minimization is not possible.

\section{Conclusions}

The $\mu$-parameter, generating the mixing between the two Higgs
doublets of the supersymmetric standard model, is generically induced
in string theories once supersymmetry is broken at low
energies. In this work, we analyzed in detail the effective
$\mu$-term, in a model-independent
way. In superstring theory there are three possible sources for
generating \nolinebreak$\mu$:
\begin{itemize}
\item[1)] The presence of a term in the K\"ahler potential which is
quadratic and analytic in the Higgs fields, but with non-analytic
moduli-dependent coefficient $H$.
\item[2)] The presence of a higher weight F-term, leading to interactions
which are not described by the standard two-derivative supergravity. In the
case of Higgs fields having non-trivial Yukawa couplings with
gauge singlet fields,
these new interactions mix with the ordinary two-derivative
couplings.
\item[3)] \hspace*{-1mm} Explicit superpotential masses, including possible
non-perturbative contributions \cite{al}.
\end{itemize}

In (2,2) compactifications, we were able to express both the K\"ahler
function $H$ and the coupling associated with the new, higher weight
interactions in terms of the moduli metric and the singlet Yukawa
couplings, generalizing special geometry to the matter field components of the
Riemann tensor. However, the resulting mass formula becomes
complicated because of the appearance of ambiguities related to the
construction of the off-shell effective action. Furthermore, Yukawa
couplings of Higgs fields with singlets produce in general a direct
superpotential mass since the singlets can aquire non-vanishing
expectation values at the scale of supersymmetry breaking. Thus, a
correct analysis requires the complete minimization of the scalar
potential which is a complicated dynamical problem.

In the case of compactifications which give rise to the particle
content of the minimal supersymmetric standard model at low energies,
there are no massless singlets coupled to Higgs particles and the above
complication does not arise. In this case, the induced $\mu$-term
depends entirely on the K\"ahler function $H$ and eventual
non-perturbative superpotential generated at the supersymmetry
breaking scale.

In (2,2) compactifications, the function $H$ satisfies a very simple
differential equation involving the moduli metrics.
As an example, we determined its dependence on
untwisted moduli in orbifold compactifications, up to the one-loop
level. We showed that it is nonvanishing only in the case in which
Higgs particles belong to the untwisted sector and are associated with a
$Z_2$-twisted internal plane. Furthermore, it depends on the moduli
of this plane only. This dependence
is consistent with large-small radius duality, with the $U$ modulus
transforming non-trivially under $SL(2,Z)$ $T$-duality, and vice
versa.

When supersymmetry breaking occurs through gaugino condensation, a
direct non-perturbative superpotential mass is generated through the
Higgs field dependence of threshold corrections to gauge couplings.
We examined these threshold corrections at the one-loop level,
for arbitrary
(2,2) compactifications. We showed that their gauge group dependent
part satisfies a simple differential equation involving only the
tree-level K\"ahler function $H$. Moreover, the gauge group
independent contribution (Green-Schwarz term) is given by the
one-loop correction to $H$, as expected. As an example, we derived
full expressions for matter-dependent threshold corrections
in orbifold models.
We also gave an effective action description of
fermion mass generation by gaugino condensation. Higgsino masses
can be expressed in terms of the gravitino mass and the
moduli vacuum expectation values.\\[1cm]
{\bf Acknowledgements}

I.A. thanks the Department of Physics at Northeastern University, and
E.G., K.S.N. and T.R.T. thank the Centre de Physique
Th\'eo\-rique at Ecole Polytechnique, for their hospitality during
completion of this work. I.A. also thanks J. Louis for useful discussions.

\newpage
\begin{figcap}
\item Feynman diagrams contributing to the effective vertex
involving two matter fermions and one auxiliary field.
\item Feynman diagrams contributing to the four-scalar amplitude
involving two matter fields and two anti-moduli.
\end{figcap}
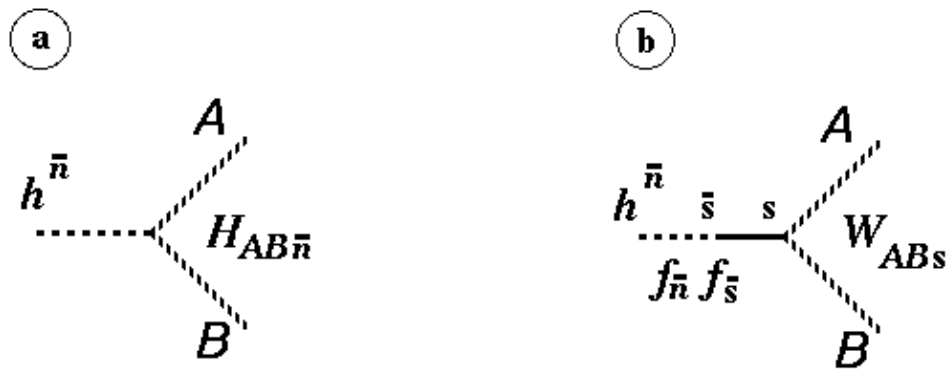
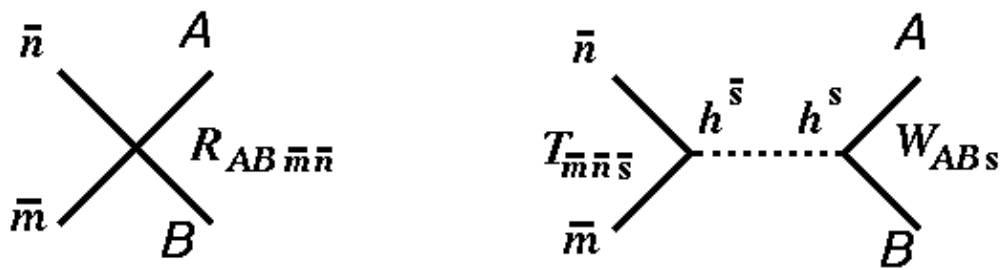
\end{document}